%% file: ISAC_inThe3GPP_NR.tex
\documentclass[twocolumn, colorlinks=true, linkcolor=blue, urlcolor=blue, citecolor=blue]{IEEEtran}   

\input{preamble.tex}
\input{Acronyms.tex}

\makeatletter

\newcommand{\vrad}{v_{\phi}}
\newcommand{\taud}{\tau_d}
\newcommand{\taus}{\tau_\textrm{R}}
\newcommand{\Ncp}{N_{\textrm{cp}}}
\newcommand{\Ts}{T_{\textrm{s}}}
\newcommand{\fc}{f_{\textrm{c}}}
\newcommand{\ns}{n_{\textrm{R}}}
\newcommand{\Kmset}{\mathcal{K}(m)}
\newcommand{\qinK}{q \in \Kmset}

\newcommand{\KcombPRS}{K_{\mathrm{comb}}^{\mathrm{PRS}}}
\newcommand{\KcombDDRS}{K_{\mathrm{comb}}^{\mathrm{DDRS}}}
\newcommand{\Mcomb}{M_{\textrm{comb}}}
\newcommand{\Nactive}{N_{\textrm{A}}}
\newcommand{\Nper}{N_{\textrm{per}}}

\begin{document}

\title{\huge{Integrated sensing and communications in the 3GPP New Radio: sensing limits}}

\author{Santiago Fernández, Javier Giménez, Mari Carmen Aguayo-Torres, José A. Cortés\\ 
Communications and Signal Processing (ComSP) Lab, Telecommunication Research Institute (TELMA), \\ Universidad de M\'alaga, E.T.S. Ingenier\'ia de Telecomunicaci\'on, Bulevar Louis Pasteur 35, 29010 M\'alaga (Spain)\\
\{sff, jgimenezdlc, mdaguayo, jca\}@uma.es}

\maketitle

\begin{abstract}
Integrated Sensing and Communications (ISAC) is regarded as a key element of the beyond–fifth-generation (5G) and sixth-generation (6G) systems, raising the question of whether current 5G New Radio (NR) signal structures can meet the sensing accuracy requirements specified by the Third Generation Partnership Project (3GPP). This paper addresses this issue by analyzing the fundamental limits of range and velocity estimation through the Cramér–Rao lower bound (CRLB) for a monostatic unmanned aerial vehicle (UAV) sensing use case currently under consideration in the 3GPP standardization process. The study focuses on standardized signals and also evaluates the potential performance gains achievable with reference signals specifically designed for sensing purposes.

The compact CRLB expressions derived in this work highlight the fundamental trade-offs between estimation accuracy and system parameters. The results further indicate that information from multiple slots must be exploited in the estimation process to attain the performance targets defined by the 3GPP. As a result, the 5G NR positioning reference signal (PRS), whose patterns may be suboptimal for velocity estimation when using single-slot resources, becomes suitable when multislot estimation is employed. Finally, we propose a two-step iterative range and radial-velocity estimator that attains the CRLB over a significantly wider range of distances than conventional maximum-likelihood (ML) estimators, for which the well-known threshold effect severely limits the distance range over which the accuracy requirements imposed by the 3GPP are satisfied. 

\end{abstract}



\section{Introduction}

Integrated Sensing and Communications (ISAC) has emerged as a key paradigm in the evolution towards beyond-\ac{5G} and \ac{6G} systems, enabling wireless networks to jointly provide communication services and environmental awareness \cite{Zhang2026}. In this context, sensing refers to the capability of extracting information about the environment—such as the presence, distance, and velocity of targets—by processing reflected radio signals. Hence, the distance and speed of a target can be estimated by measuring the delay and the Doppler shift of the line-of-sight echo between the transmitter and the object \cite{Andersson2021}. This differs from the positioning functionality currently defined in \ac{5G}, where the \ac{BS} transmits a reference signal that is used by the \ac{UE} to estimate its position (which is reported back to the \ac{BS}). The sensing functionality inherently combines detection and parameter estimation, and its performance is typically characterized in terms of both reliability and estimation accuracy. 

The integration of sensing into cellular systems enables a wide range of applications \cite{Li2023}\cite{Cui2025}. On the one hand, sensing as a service allows the network to provide environmental information to external systems, such as autonomous vehicles or smart infrastructure. On the other hand, sensing can be leveraged to improve network operation itself, enabling functionalities such as blockage detection, beam management, and proactive resource allocation. However, these benefits come at the cost of a fundamental trade-off between sensing performance and resource utilization, since sensing requires the allocation of time-frequency resources that would otherwise be used for communication.

From a system perspective, sensing can be implemented under two different operational modes: monostatic and bistatic \cite{Andersson2021}. In the former, the signal is received and analyzed by the same \ac{BS}, while in the bistatic case the signal is transmitted by a \ac{BS} and received and analyzed by a different one. Monostatic operation simplifies synchronization and geometry but requires in-band full-duplex capabilities with stringent self-interference cancellation \cite{Zhang2022}. Bistatic (or multi-static, in case the signal is received by more than one \ac{BS}) configurations relax this requirement but introduce additional complexity in terms of synchronization, coordination, and error propagation across distributed nodes.

Within the \ac{3GPP} framework, sensing is being actively investigated as part of the evolution towards \ac{6G}, of which sensing is considered a \textit{day-zero functionality}. The initial study item has defined a use case consisting on \ac{UAV} sensing using a monostatic configuration \cite{3gpp_R1_2507427_2025}. It has defined detection metrics, such as missed detection and false alarm probabilities, as well as estimation accuracy metrics for parameters such as target distance and velocity, which must remain below predefined thresholds with a 90\% confidence level. 

A key open question is whether existing \ac{5G} NR signal structures are fundamentally capable of meeting these sensing accuracy requirements. Since the achievable estimation performance is inherently limited by the structure of the transmitted signals and the allocated resources, it is essential to assess these limits from a theoretical perspective before considering specific estimation algorithms. 

\begin{figure*}[b]
\begin{equation}
\begin{aligned}
    &y_{\textrm{\textrm{LP}}}(n)=\sqrt{\alpha_{\textrm{T}}} x_{\textrm{\textrm{LP}}}(n-\tau(n\Ts)/\Ts)e^{-j(2\pi\fc \tau(n\Ts)-\phi_S)}+w_{\textrm{\textrm{LP}}}(n)=\\&\frac{\sqrt{\alpha_{\textrm{T}}}}{N}   \sum_{m} \sum_{k\in \mathcal{K}(m)} \!\!\!\!X_{k,m} e^{j\frac{2\pi k}{N} \left(n - \frac{\taud}{\Ts} - \frac{2 \vrad n}{c_0}  - mL - \Ncp\right)} e^{-j\left( 2 \pi \fc \left( \taud + \frac{2 \vrad n \Ts}{c_0} \right)-\phi_S\right)} \omega_{\textrm{TX}}\left( \Ts \left( n - mL- \frac{\taud}{\Ts} - \frac{2 \vrad n }{c_0}  \right) \right) + w_{\textrm{\textrm{LP}}}(n).
\end{aligned}
\label{eq:SignalModel:LP_DT_received_signal}
\end{equation}
\end{figure*}

The \ac{CRLB} provides a fundamental benchmark for range and velocity estimation. While widely employed in \ac{ISAC} applications \cite{Bekkerman2006,Gaudio2019,Liu2022,Wei_2023,Soltani2025}, existing expressions consider neither the specific structure of \ac{5G} \ac{NR} signals nor the studied use case \cite{Bekkerman2006,Gaudio2019,Soltani2025}, rely on channel models that yield overly complex expressions evaluable only numerically \cite{Soltani2025}, or model the Doppler as a frequency shift rather than as the actual spectral compression/expansion \cite{Wei_2023}. 

In this work, we derive the \ac{CRLB} for the \ac{UAV} distance and velocity estimation under different \ac{5G} NR-based grid configurations, including standardized \ac{PRS} patterns and novel sensing-oriented designs. The presented \ac{CRLB} values are employed to compute the highest accuracy that can be obtained in the estimation of the range and velocity, which are compared to the limits currently agreed in \cite{3gpp_R1_2509243_2025}. This provides insights into the fundamental trade-offs between estimation accuracy and resource utilization, and identifies the conditions under which the target sensing performance can be achieved. Furthermore, we propose a two-step iterative range and velocity estimation algorithm which attains the \ac{CRLB} in a much wider range of \ac{SNR} than plain \ac{ML} approaches, whose threshold effect made them inadequate to comply with the imposed requirements. 

The rest of this paper is organized as follows. Section \ref{Sec:SystemModel} describes the system model for the considered use case. Then, the \acp{CRLB} for velocity and range estimations are derived in Section \ref{Sec:CRLB}. The different configurations for the sensing pattern are summarized in Section \ref{Sec:SensingPatterns}. In Section \ref{Sec:ML_estimators}, the proposed two-stage iterative range and velocity estimation algorithm is presented. Numerical results and discussion are presented in Section \ref{Sec:NumericalResults&Discussion}. Finally, conclusions are outlined in Section \ref{Sec:Conclusion}.

\section{System model}
\label{Sec:SystemModel}
\begin{figure}[!ht]
    \centering
    \includegraphics[width=0.95\linewidth]{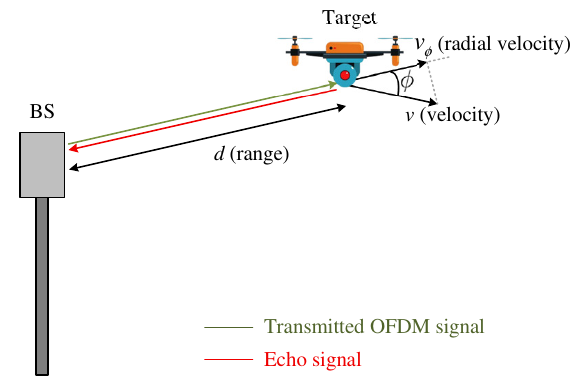}
    \caption{Sensing use case consisting of a monostatic configuration and a single \ac{UAV}.}
    \label{fig:System_model}
\end{figure}

The considered use case consists of a monostatic configuration for \ac{UAV} sensing, according to \cite{3gpp_R1_2507427_2025}. Since we are interested in obtaining an upper bound of the achievable sensing performance, the case where only the target whose range and speed is to be estimated is present, as illustrated in Fig. \ref{fig:System_model}.

The discrete-time low-pass equivalent expression of an \ac{OFDM} signal with $N$ subcarriers and $N_{\textrm{cp}}$ samples of cyclic prefix can be expressed as
\begin{equation}
\begin{aligned}
    x_{\textrm{\textrm{LP}}}(n) &= \frac{1}{N} \sum_{m} \sum_{k \in \mathcal{K}(m)} X_{k,m} e^{j \frac{2\pi}{N} k(n - N_{cp} - mL)}\\ &\times \omega_{\textrm{\textrm{TX}}}(n - mL),
\end{aligned}
    \label{eq:SignalModel:LP_DT_transmitted_signal}
\end{equation}
where $X_{k,m}$ is the constellation symbol carried by the $k$-th subcarrier in the $m$-th symbol, $L=N+N_{\mathrm{cp}}$ is the number of samples per \ac{OFDM} symbol and $w_{\mathrm{TX}}\left(n\right)$ is a rectangular window of length $L$. The set of active carriers used in the $m$-th \ac{OFDM} symbol is denoted as  $\mathcal{K}(m)$. 

The signal $x_{\textrm{\textrm{LP}}}(t)$ is then converted from digital to analog form, frequency-shifted to $\fc$ and converted to a real one
\begin{equation}
x(t)=\sqrt{2}\Re\{x_{\textrm{\textrm{LP}}}(t)e^{j2\pi\fc t}\},
\end{equation}
where $\Re\{\cdot\}$ denotes the real part.

The signal $x(t)$ is then reflected in the single \ac{ST} located $d$ meters away from the \ac{BS} and moving at a relative radial-velocity $\vrad=v\cos(\phi)$. Let us denote the scattering amplitude of the \ac{ST} as ${S=|S|e^{j\phi_S}}$, which is related to the radar cross section as $\sigma_{\textrm{RCS}}=|S|^2$, and that is assumed to be independent of the frequency and azimuth and elevation angles. Assuming free-space propagation, the attenuation given by the radar equation is $\alpha_{\mathrm{T}}=\frac{c_0^2\sigma_{\textrm{RCS}}}{(4\pi)^3 d^4\,\fc^2}$, and the received echo can be expressed as
\begin{equation}
y(t)=\sqrt{2} \Re\bigg\{\sqrt{\alpha_{\textrm{T}}} x_{\textrm{\textrm{LP}}}(t-\tau(t))e^{-j(2\pi\fc \tau(t)-\phi_S)}e^{j2\pi\fc t}\bigg\}+w(t),
\end{equation}
where $w(t)$ is the noise at the receiver and ${\tau(t)=2d/c_0+2\vrad t/c_0}$. For simplicity of notation, $\tau_d=2d/c_0$ is defined. 

Since the transmitter and the receiver are co-located, perfect carrier and sampling synchronization can be assumed. Denoting the sampling period by $\Ts = \frac{1}{N \Delta f}$, the discrete-time low-pass equivalent of the echo received at the \ac{BS} can be written as in (\ref{eq:SignalModel:LP_DT_received_signal}). 

In order to compensate for part of the echo delay, the receiver is assumed to shift the \ac{DFT} window by $\taus=\ns\Ts$ seconds relative to the start of the transmitted symbol. Disregarding the \ac{ICI} caused by the Doppler effect and assuming that the propagation delay is smaller than the cyclic prefix duration, the constellation symbols received at subcarrier $k$ in the $m$-th OFDM symbol are then given by
\begin{equation}
    Y_{q,m} = \sqrt{\alpha_{\textrm{T}}} X_{q,m}  e^{-j 2\pi \varphi_{q,m}(\taud, \vrad)} + W_{q,m}, 
    \label{eq:SignalModel:Y_qm_post_FFT}
\end{equation}
where the noise sample in the $q$-th subcarrier of the $m$-th OFDM symbol is modeled as $W_{q,m}\sim CN\left(0,\sigma_w^2\right)$, and 
\begin{equation}
\begin{aligned}
    \varphi_{q,m}(\taud, \vrad) &=  \underbrace{\fc \taud -\frac{\phi_S}{2\pi}}_{\psi} + \Delta f  q \left( \taud -\taus \right)\\  &+ (\fc + \Delta f  q)\frac{2\vrad}{c_0}\delta_m\Ts,
\end{aligned}
    \label{eq:SignalModel:Phase_of_Y_qm}
\end{equation}
where $\Delta f$ is the \ac{SCS} and ${\delta_m = \left( n_{\textrm{R}} + N_{\rm cp} + \frac{N-1}{2} + mL \right)}$.

The phase in (\ref{eq:SignalModel:Phase_of_Y_qm}) consists of three terms: two that depend on $\taud$ and one that depends on $\vrad$. Our goal is to estimate the distance to the \ac{ST} and its radial-velocity from the information embedded in this phase term. In practice, no information about $\taud$ can be extracted from the phase term $\fc \taud$ because the product $\fc~\taud~\gg~1$, which causes phase ambiguity. Accordingly, both $\fc\taud$ and $\phi_S$ are grouped in the unknown parameter $\psi$.

It is worth to highlight that expression (\ref{eq:SignalModel:Phase_of_Y_qm}) differs from the widely used one given in \cite{Sturm_2011}. In the latter, the phase shift associated to the radial speed, $\vrad$, is independent of $\Delta f q$. This is due to the more accurate modeling of the Doppler effect used in (\ref{eq:SignalModel:LP_DT_received_signal}), which causes a compression or expansion of the signal in addition to the frequency shift considered in \cite{Sturm_2011}.

\section{Cramer-Rao lower bound of range and velocity}
\label{Sec:CRLB}
Let us now determine the \ac{CRLB} of the range and radial-velocity estimation from a set of \ac{OFDM} symbols with indexes $m \in \mathcal{M}$, with $M=|\mathcal{M}|$. To this end, it can be firstly observed that $Y_{q,m}\sim CN\left(\sqrt{\alpha_{\textrm{T}}} X_{q,m}  e^{-j 2\pi \varphi_{q,m}(\taud, \vrad)},\sigma_w^2\right)$. Since noise samples in different subcarriers and \ac{OFDM} symbols are independent, the \ac{LLF} can be expressed as in (\ref{eq:loglikelihood_PRS}) and the Fisher information as in (\ref{eq:CRLB:Fished_Information_Matrix_definition}), where it has been assumed that the constellation symbols of the sensing signals have equal amplitude, $|X_{q,m}|^2=E_X$. Denoting the \ac{SNR} in the $q$-th subcarrier as ${\textrm{SNR}_q=\alpha_{\textrm{T}}E_X/\sigma_w^2}$, the elements $I_{ij} = \left[ \mathbf{I}(\taud,\vrad,\psi) \right]_{i,j}$ are given by

\begin{equation}
\def\qinK{q\in\Kmset}
\def\dmn{\left( n_s + N_{\rm cp} + \frac{N-1}{2} + m'L + nN_{\rm symb}^{\rm ts} L \right)}
    \begin{aligned}
    &I_{11} = \sum_{m\in\mathcal{M}} \sum_{\qinK} \textrm{SNR}_q (\Delta f q)^2\\
    &I_{12} = I_{21} = \frac{2T_s}{c_0}\sum_{m\in\mathcal{M}} \sum_{\qinK} \textrm{SNR}_q \Delta f q (f_c + \Delta f q) \delta_m\\
    &I_{13} = I_{31} = \sum_{m\in\mathcal{M}} \sum_{\qinK} \textrm{SNR}_q \Delta f q\\
    &I_{22} = \left(\frac{2T_s}{c_0}\right)^{\!\!2} \sum_{m\in\mathcal{M}} \sum_{\qinK} \textrm{SNR}_q (f_c + \Delta f q)^2 \delta_m^2\\
    &I_{23} = I_{32} = \frac{2T_s}{c_0}\sum_{m\in\mathcal{M}} \sum_{\qinK} \textrm{SNR}_q (f_c + \Delta f q) \delta_m,\\
    &I_{33} = \sum_{m\in\mathcal{M}} \sum_{\qinK} \textrm{SNR}_q. \\
    \end{aligned}
    \label{eq:CRLB:Fisher_info_matrix_terms}
\end{equation}

The \ac{CRLB} for the estimation of $d$ and $\vrad$ can then be expressed as 
\begin{equation}
    \begin{aligned}
        \textrm{VAR}(\hat{d}) &\geq \frac{c_0^2}{32\pi^2} \frac{I_{22}I_{33} - I_{23}^2}{\textrm{det}(\mathbf{I}(\taud,\vrad,\psi))}\\
        \textrm{VAR}(\hat{\vrad}) &\geq \frac{1}{8\pi^2} \frac{ I_{11}I_{33} - I_{13}^2}{\textrm{det}(\mathbf{I}(\taud,\vrad,\psi))}
    \end{aligned},
    \label{eq:CRLB:CLRB_simple_exp}
\end{equation}
where $\textrm{det}\left(\cdot\right)$ denotes the determinant operator.

\begin{figure*}[t!]
\begin{equation}
\begin{aligned}
    \Lambda(\tau_d, v_{\phi}, \psi)  =- \sum_{m\in\mathcal{M}} \sum_{\qinK} \!\!\!\!\! \Big( \ln\left(\pi\sigma_w^2\right) + \frac{1}{\sigma_w^2} \left(|Y_{q,m}|^2 +\alpha_{\textrm{T}}E_X\right)\Big) + \frac{2\sqrt{\alpha_{\textrm{T}}}}{\sigma_w^2} \sum_{m\in\mathcal{M}} \sum_{\qinK} \!\!\! \Re\left\{X_{q,m}e^{-j2\pi \varphi_{q,m}(\taud, \vrad)} Y_{q,m}^* \right\}.
\end{aligned}
    \label{eq:loglikelihood_PRS}
\end{equation}

\begin{equation}
    \begin{aligned}
    &\mathbf{I}(\taud, \vrad , \psi) = 
    & \left[\begin{array}{ccc}
       -\mathbb{E} \left[ \frac{\partial^2 \Lambda(\tau_d, v_{\phi}, \psi)}{\partial \tau_d^2}  \right]
     & -\mathbb{E} \left[ \frac{\partial^2 \Lambda(\tau_d, v_{\phi}, \psi)}{\partial \tau_d \partial  v_{\phi}}  \right] & -\mathbb{E} \left[ \frac{\partial^2 \Lambda(\tau_d, v_{\phi}, \psi)}{\partial \tau_d \partial  \psi}  \right] \\
      -\mathbb{E} \left[ \frac{\partial^2 \Lambda(\tau_d, v_{\phi}, \psi)}{\partial v_{\phi} \partial \tau_d}  \right] & -\mathbb{E} \left[ \frac{\partial^2 \Lambda(\tau_d, v_{\phi}, \psi)}{\partial v_{\phi}^2}  \right] & -\mathbb{E} \left[ \frac{\partial^2 \Lambda(\tau_d, v_{\phi}, \psi)}{\partial v_{\phi} \partial  \psi}  \right] \\
      -\mathbb{E} \left[ \frac{\partial^2 \Lambda(\tau_d, v_{\phi}, \psi)}{\partial  \psi \partial \tau_d }  \right] & -\mathbb{E} \left[ \frac{\partial^2 \Lambda(\tau_d, v_{\phi}, \psi)}{\partial  \psi \partial v_{\phi} }  \right] & -\mathbb{E} \left[\frac{\partial^2 \Lambda(\tau_d, v_{\phi}, \psi)}{\partial  \psi^2 } \right]   
    \end{array}    \right]
    \end{aligned}.
    \label{eq:CRLB:Fished_Information_Matrix_definition}
\end{equation}
\end{figure*}

Assuming that all subcarriers experience the same \ac{SNR}, $\textrm{SNR}_q=\textrm{SNR}$, that $\fc \gg \Delta f q$, and that all symbols use the whole set of active subcarriers for sensing purposes, $\Kmset=\mathcal{K}$, the \ac{CRLB} can be compactly written as

\begin{equation}
    \def\sumq{\sum_q \!q}
    \def\sumqs{\sum_q \!q^2}
    \def\sumdm{\sum_m \!\delta_m}
    \def\sumdms{\sum_m \!\delta_m^2}
    \begin{aligned}
        &\textrm{VAR}(\hat{d}) \geq \Gamma\cdot \frac{1}{\Delta f^2}\cdot \frac{1}{\frac{1}{\Nactive} \sumqs - (\frac{1}{\Nactive}\sumq)^2}\\
        &\textrm{VAR}(\hat{v_{\phi}}) \geq \Gamma \cdot \frac{N^2 \cdot \Delta f^2}{f_c} \cdot\frac{1}{\frac{1}{M}\sumdms - (\frac{1}{M}\sumdm)^2}
    \end{aligned},    
    \label{eq:CRLB:CRLB_full_slot_compact}
\end{equation}
with 
\begin{equation}
\Gamma = \frac{c_0^2}{32 \pi^2 \cdot \textrm{SNR} \cdot M \cdot \Nactive}.
\end{equation}

As expected, the \ac{CRLB} of both range and radial-velocity improves with the \ac{SNR} and the number of observations $M \cdot \Nactive$. However, while increasing the \ac{SCS}, $\Delta f$, improves the range estimation, it degrades the radial-velocity one. Increasing the carrier frequency improves the velocity estimation without affecting the range one. Interestingly, range and radial-velocity estimation improve with increasing subcarrier and symbol index variance, respectively, consistent with \cite{Wang2024}.

The \acp{CRLB} in (\ref{eq:CRLB:CRLB_full_slot_compact}) express the upper bound of the estimation process in terms of the variance. However, \ac{3GPP} has agreed to use accuracy for a given confidence level as \ac{KPI} \cite{3gpp_R1_2507427_2025}. Both magnitudes can be related as follows. The accuracy of an estimator $\hat{\theta}$ for a confidence level $(1-\alpha)$ ($\alpha$ is the significance level), is defined as the value of $\Delta\theta$ that fulfills

\begin{equation}
\textrm{P}_r\left(|\hat{\theta}-\theta|\leq\Delta\theta\right)=1-\alpha. 
\label{eq:Pr_accuracy}
\end{equation}

Assuming that $\hat{\theta}$ is normally distributed with variance and bias $\textrm{VAR}(\hat{\theta})$ and $\textrm{B}(\hat{\theta})$, respectively\footnote{While the \ac{CRLB} is derived for unbiased estimators, we herein provide the relation for the general case of a biased estimator.}, (\ref{eq:Pr_accuracy}) can be expressed in terms of the Gaussian Q-function, $Q(\cdot)$, as
\begin{equation}
1-\alpha = 1 - Q\left( \frac{\Delta\theta - \textrm{B}(\hat{\theta})}{\sqrt{\textrm{VAR}(\hat{\theta})}}\right) - Q\left( \frac{\Delta\theta + \textrm{B}(\hat{\theta})}{\sqrt{\textrm{VAR}(\hat{\theta})}}\right),
    \label{eq:CRLB:Accuracy_expression}
\end{equation}
from which $\Delta\theta$ can be obtained. 

\section{Sensing patterns}
\label{Sec:SensingPatterns}

\subsection{Full-slot}
The upper bound in the sensing performance is attained when all active subcarriers of all the \ac{OFDM} symbols in the slot are employed for this purpose. This signal pattern will be hereafter referred to as full-slot. The number of active subcarriers, $N_{\mathrm{A}}$, is smaller than the total number of subcarriers ($N_{\mathrm{A}} < N$), as defined in~\cite{3gpp38.104}, where the maximum number of \acp{RB} is determined by the signal bandwidth. The set of subcarrier indices is then given by 
\begin{equation}
    \Kmset= \mathcal{K}=\bigg\{-\frac{N_{\mathrm{A}}}{2},\ldots, \frac{N_{\mathrm{A}}}{2}-1\bigg\}.   
\end{equation}

\subsection{Positioning reference signal (PRS)}
The \ac{5G} \ac{NR} standard defines a set of downlink reference signals for different purposes. Among the most representative are the \ac{SS}, which enables the \ac{UE} to detect a cell; the \ac{DM-RS}, used for downlink channel estimation; the \ac{CSI-RS}, used to assess link quality; and the \ac{PRS}, which allows the \ac{UE} to estimate time-of-arrival, typically using \acp{PRS} from multiple \acp{BS}~\cite{ETSI38.211,Dahlman2018}. Among the reference signals, the \ac{PRS} has important advantages for downlink sensing purposes, such as higher flexibility and density (as it admits multiple configurations) and being an always-on signal. 

The mapping of PRS symbols onto the physical resource grid is primarily governed by the \textit{comb size}, denoted as $K^{\mathrm{PRS}}_{\mathrm{comb}}~\in~\{2,\,4,\,6,\,12\}$, which specifies the spacing between subcarriers allocated to \ac{PRS} within each \ac{RB}. For instance, $K^{\mathrm{PRS}}_{\mathrm{comb}} = 4$ implies that one out of every four subcarriers within an \ac{OFDM} symbol carries \ac{PRS}.

The corresponding time--frequency indices of the \ac{PRS} pattern $\{q,m\}$ are defined as
\begin{equation}
\begin{aligned}
    m &= m_{\mathrm{start}}^{\mathrm{PRS}}, \dots, m_{\mathrm{start}}^{\mathrm{PRS}} + M_{\mathrm{PRS}} - 1, \\
    q &= l \cdot K_{\mathrm{comb}}^{\mathrm{PRS}} + \big( (q_{\mathrm{offset}}^{\mathrm{PRS}} + q') \bmod K_{\mathrm{comb}}^{\mathrm{PRS}} \big), \\
    l &= 0, 1, \dots, L_{\mathrm{PRS}} - 1,
\end{aligned}
\label{eq:k_m_indices_definition_PRS}
\end{equation}
where $M_{\mathrm{PRS}} \in \{1, \, 2, \, 4, \, 6, \, 12\}$ denotes the number of OFDM symbols within a slot allocated to \ac{PRS}, $q_{\mathrm{offset}}^{\mathrm{PRS}}~\in~\{0, \dots, K_{\mathrm{comb}}^{\mathrm{PRS}}-1\}$ is the frequency offset, and $L_{\mathrm{PRS}} = {N_{\mathrm{A}}}/{K_{\mathrm{comb}}^{\mathrm{PRS}}}$ is the number of \ac{PRS} repetitions across frequency. Only the following combinations of $\{M_{\mathrm{PRS}}, K_{\mathrm{comb}}^{\mathrm{PRS}}\}$ are supported in the specification,  
\begin{equation}
    \begin{aligned}
\{ M_{\rm PRS}, \KcombPRS \} \in \big\{ &\{1,2\}, \{2,2\}, \{4,2\}, \{6,2\}, \{12,2\}, \\ &\{1,4\}, \{4,4\}, \{12,4\}, \{1,6\}, \{6,6\},\\ &\{12,6\}, \{1,12\}, \{12,12\} \big\}   
    \end{aligned}.
\end{equation}
However, in order to keep the overhead constant among configurations, only the cases in which $\KcombPRS  = M_{\rm PRS}$ are employed.

The index $q'$ in (\ref{eq:k_m_indices_definition_PRS}) depends on the relative symbol index $(m~-~m_{\mathrm{start}}^{\mathrm{PRS}})$ and $K_{\mathrm{comb}}$, as summarized in Table~\ref{table:q_prime_index_per_m}.

\begin{table}[ht]
    \centering
    \caption{Value of $q'$ for a given ($m - m_{\rm start}^{\rm PRS}$) and $\KcombPRS$}
    \begin{tabular}{c|c|c|c|c|c|c|c|c|c|c|c|c}
       \multirow{2}{*}{$\KcombPRS$}  & \multicolumn{12}{c}{\parbox{7 cm}{\centering Relative index of the $m$-th symbol within the time slot ($m - m_{\rm start}^{\rm PRS}$)}} \\ & 0 & 1 & 2 & 3 & 4 & 5 & 6 & 7 & 8 & 9 & 10 & 11\\ \hline
       2 & 0 & 1 & 0 & 1 & 0 & 1 & 0 & 1 & 0 & 1 & 0 & 1 \\
       4 & 0 & 2 & 1 & 3 & 0 & 2 & 1 & 3 & 0 & 2 & 1 & 3 \\
       6 & 0 & 3 & 1 & 4 & 2 & 5 & 0 & 3 & 1 & 4 & 2 & 5 \\
       12 & 0 & 6 & 3 & 9 & 1 & 7 & 4 & 10 & 2 & 8 & 5 & 11 
    \end{tabular}
    \label{table:q_prime_index_per_m}
\end{table}

\begin{figure}[ht!]
\centering
\begin{subfigure}[t]{0.24\textwidth}
  \centering
  \includegraphics[width=\linewidth]{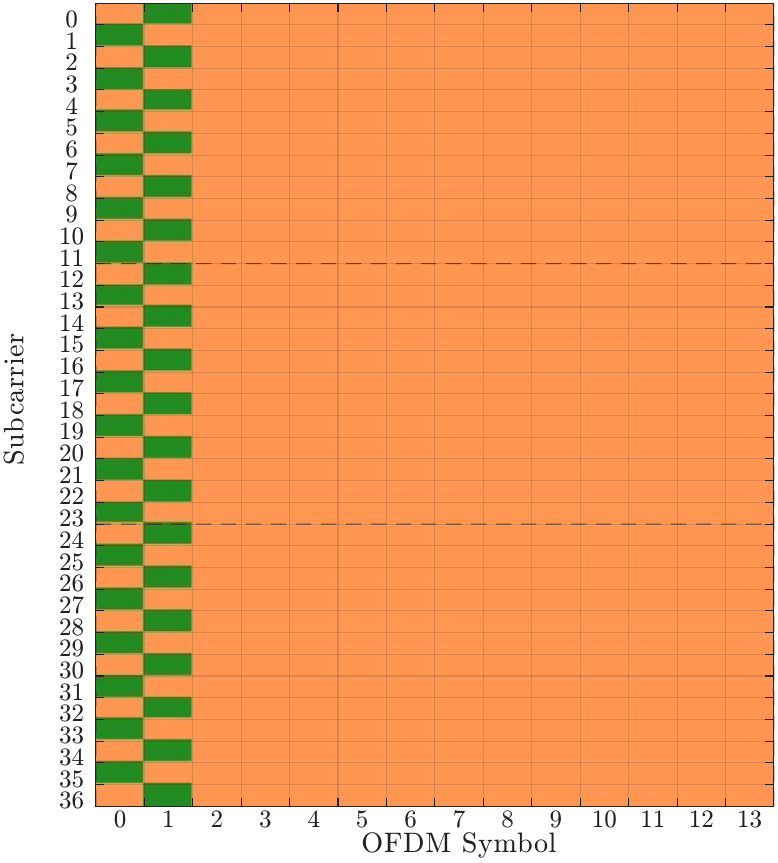}
  \caption{$K_{\rm comb}^{\rm PRS}=2$.\vspace{0.03cm}}
  \label{fig:GRID_Comb2_1Slot_PRS} 
  
\end{subfigure}
\hfill
\begin{subfigure}[t]{0.24\textwidth}
  \centering
  \includegraphics[width=\linewidth]{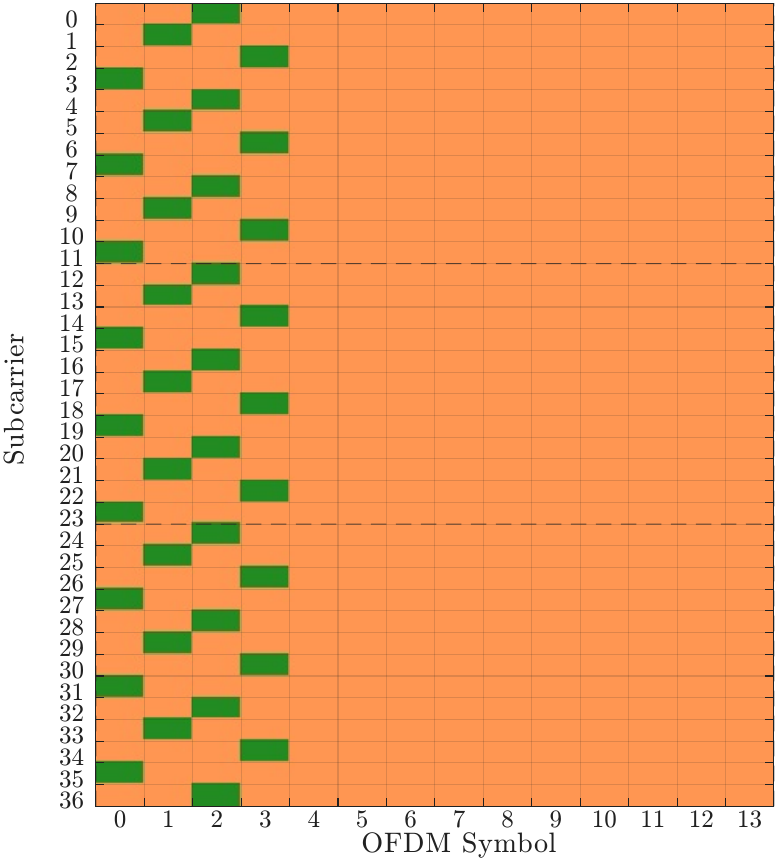}
  \caption{$K_{\rm comb}^{\rm PRS}=4$.}
  \label{fig:GRID_Comb4_1Slot_PRS}  
  
\end{subfigure}
\hfill
\begin{subfigure}[t]{0.24\textwidth}
  \centering
  \includegraphics[width=\linewidth]{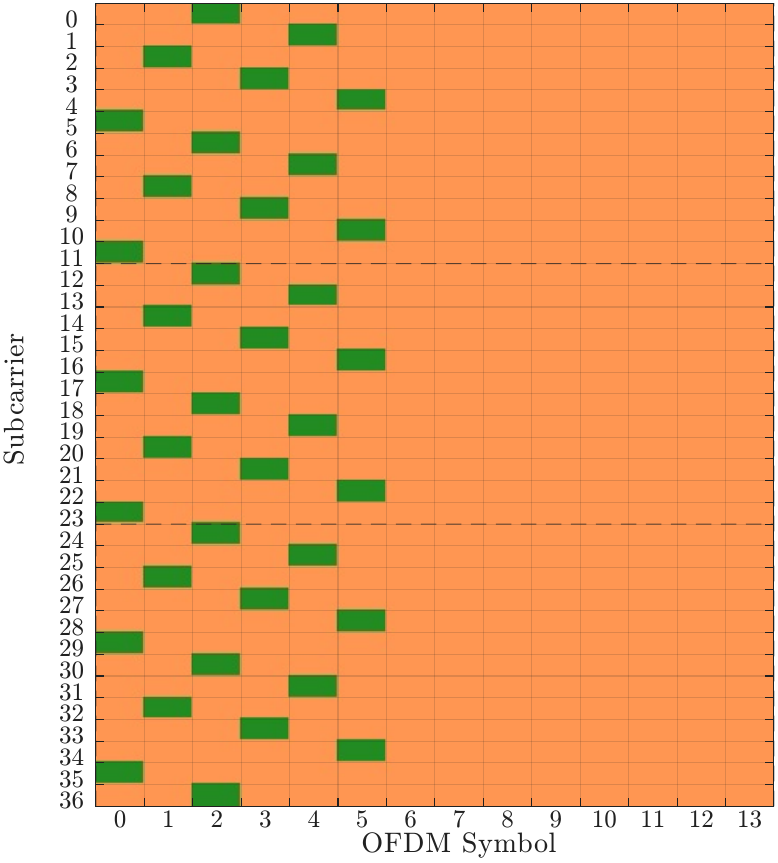}
  \caption{$K_{\rm comb}^{\rm PRS}=6$.}
  \label{fig:GRID_Comb6_1Slot_PRS}  
  
\end{subfigure}
\hfill
\begin{subfigure}[t]{0.24\textwidth}
  \centering
  \includegraphics[width=\linewidth]{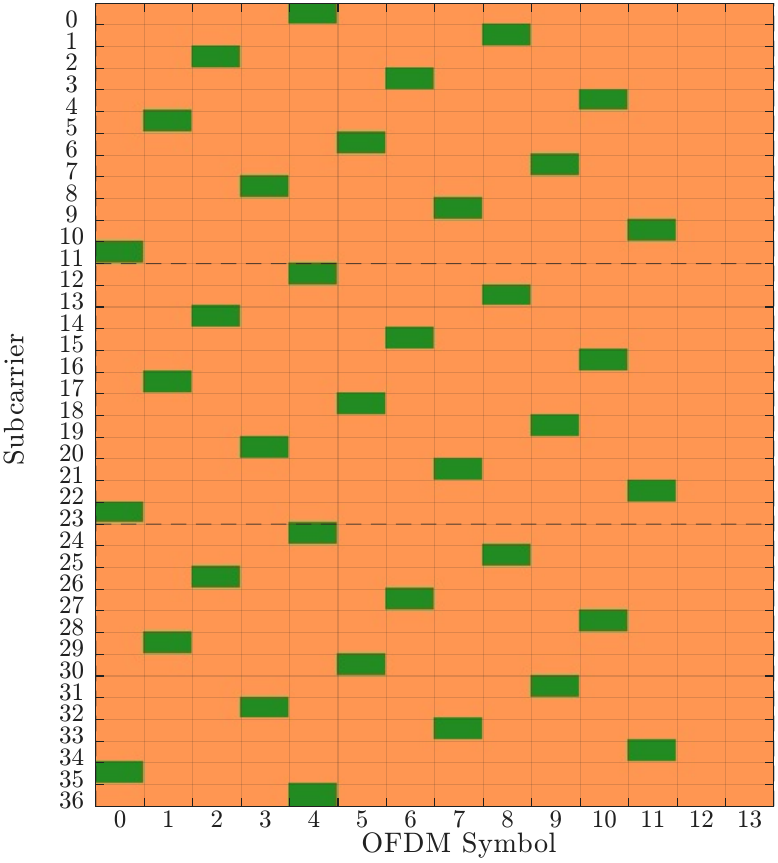}
  \caption{$K_{\rm comb}^{\rm PRS}=12$.}
  \label{fig:GRID_Comb12_1Slot_PRS}  
  
\end{subfigure}
\caption{Example of \ac{PRS} patterns with the same overhead. The green resource elements are the ones for sensing, the orange ones are used for other purposes.}
\label{fig:GRID_Comb_2_4_6_12_1Slot_PRS}
\end{figure}

For illustrative purposes, Fig. \ref{fig:GRID_Comb_2_4_6_12_1Slot_PRS} depicts example patterns corresponding to the four possible comb sizes of the \ac{PRS} within a slot. While only three \acp{RB} are shown, the \ac{PRS} pattern covers the whole set of active carriers. For simplicity, it has been assumed that $m_{\mathrm{start}}^{\mathrm{PRS}} = 0$ and $q_{\mathrm{offset}}^{\mathrm{PRS}} = 0$. Under these assumptions, the indices in (\ref{eq:k_m_indices_definition_PRS}) reduce to
\begin{equation}
\begin{aligned}
    m &= 0, 1, \dots, M_{\mathrm{PRS}} - 1, \\
    q &= q' + l \cdot \KcombPRS - \frac{N_{\mathrm{A}}}{2}, \\
    l &= 0, 1, \dots, L_{\mathrm{PRS}} - 1,
\end{aligned}
\label{eq:k_m_indices_definition_PRS_bis}
\end{equation}
where the additional term $-\frac{N_{\mathrm{A}}}{2}$ centers the subcarrier indices around zero. As seen, each \ac{OFDM} symbol $m$ is associated with a distinct subset of \ac{PRS} subcarriers, $\Kmset$.

\subsection{DDRS}
Since \ac{PRS} signals where not designed for sensing purposes, they have some drawbacks when used for this application. For instance, each carrier is used only once in each time slot, which hinders speed estimation. Similarly, for $\KcombPRS=2$, the two symbols used in each time slot are consecutive, which also yields poor performance when speed is estimated using the \ac{PRS} resources of a single slot. This can be explained in terms of the \ac{CRLB} in (\ref{eq:CRLB:CRLB_full_slot_compact}), where the variance of the symbol indices used in the estimation is extremely small. 

To assess whether better sensing performance may be obtained with specially designed patterns, we here propose a new reference signal, which will be referred to as \ac{DDRS}, to which the following design constraint are imposed. First, it must have a regular grid, using the same \acp{RE} in all \ac{OFDM} symbols. Hence, the employed  frequency and time comb sizes, $\KcombDDRS$ and $M_{\mathrm{comb}}^{\mathrm{DDRS}}$ respectively, differ from those used in \ac{PRS}. Second, the spacing between symbols with \ac{DDRS} resources should be equal (or almost) spaced. Finally, \ac{DDRS} must have the same overhead as the \ac{PRS}. 

Denoting the number of \ac{OFDM} symbols per time slot as $M_{\mathrm{OFDM}}^{\mathrm{TS}}$, the time--frequency indices of the \ac{DDRS} patterns $\{q,m\}$ are given by
\begin{equation}
\begin{aligned}
    m &= m_{\mathrm{start}}^{\mathrm{DDRS}} + \ell M_{\rm comb}^{\rm DDRS}, \\
    \ell &= 0, \,1, \, \ldots \, , \, M_{\rm DDRS} - 1 ,\\
    q &= l \cdot \KcombDDRS, \\
    l &= 0, 1, \dots, L_{\mathrm{DDRS}} - 1,
\end{aligned}
\label{eq:k_m_indices_definition_DDRS}
\end{equation}
where $M_{\rm comb}^{\rm DDRS} = \left\lfloor \dfrac{M_{\mathrm{OFDM}}^{\mathrm{TS}}}{\KcombDDRS} \right\rfloor$, with $\left\lfloor \cdot \right\rfloor$ corresponding to the floor operation,  $L_{\mathrm{DDRS}} = {N_{\mathrm{A}}}/{\KcombDDRS}$ is the number of \ac{DDRS} repetitions across frequency and $M_{\mathrm{DDRS}} \in \{2,\, 4, \, 6, \, 7, \, 12, \, 14\}$ denotes the number of OFDM symbols within a slot allocated to \ac{DDRS}. 

In order to keep the overhead constant and equal to the \ac{PRS} cases, the following configurations, denoted as $\{M_{\mathrm{DDRS}}, \KcombDDRS\}$, are supported,  
\begin{equation}
    \begin{aligned}
\{ M_{\rm DDRS}, \KcombDDRS \} \in \big\{ & \{2,2\},  \{4,4\}, \{6,6\}, \\ & \{7,7\} , \{12,12\}, \{14,14\} \big\},   
    \end{aligned}
\end{equation}

\begin{figure}[ht!]
\centering
\begin{subfigure}[t]{0.24\textwidth}
  \centering
  \includegraphics[width=\linewidth]{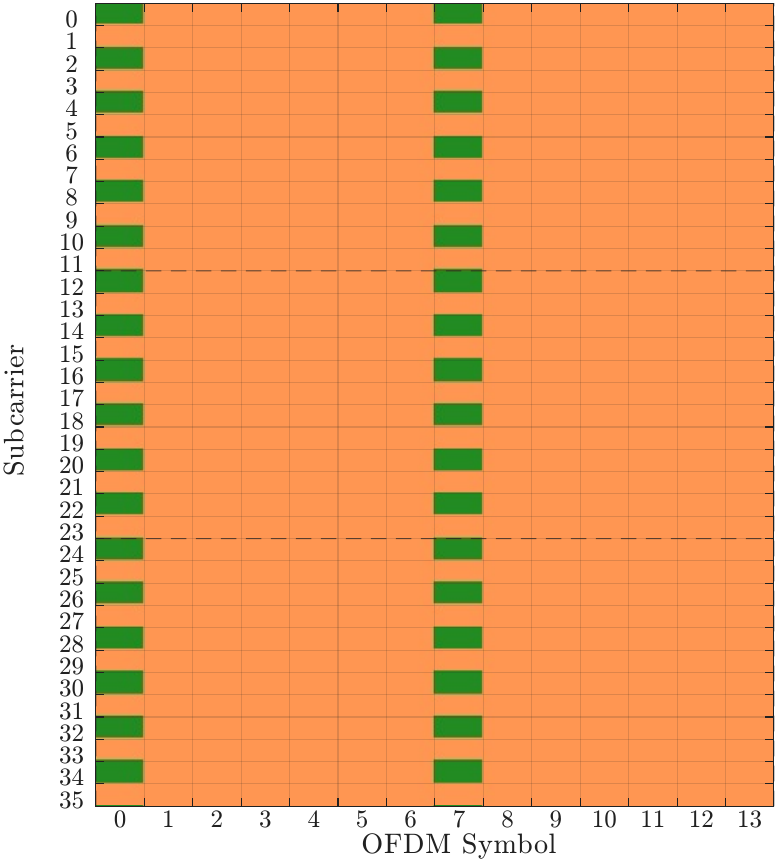}
  \caption{$\{ M_{\rm DDRS}, \KcombDDRS \} =   \{2,2\}$.\vspace{0.03cm}}
  \label{fig:GRID_Comb2_1Slot} 
  
\end{subfigure}
\hfill
\begin{subfigure}[t]{0.24\textwidth}
  \centering
  \includegraphics[width=\linewidth]{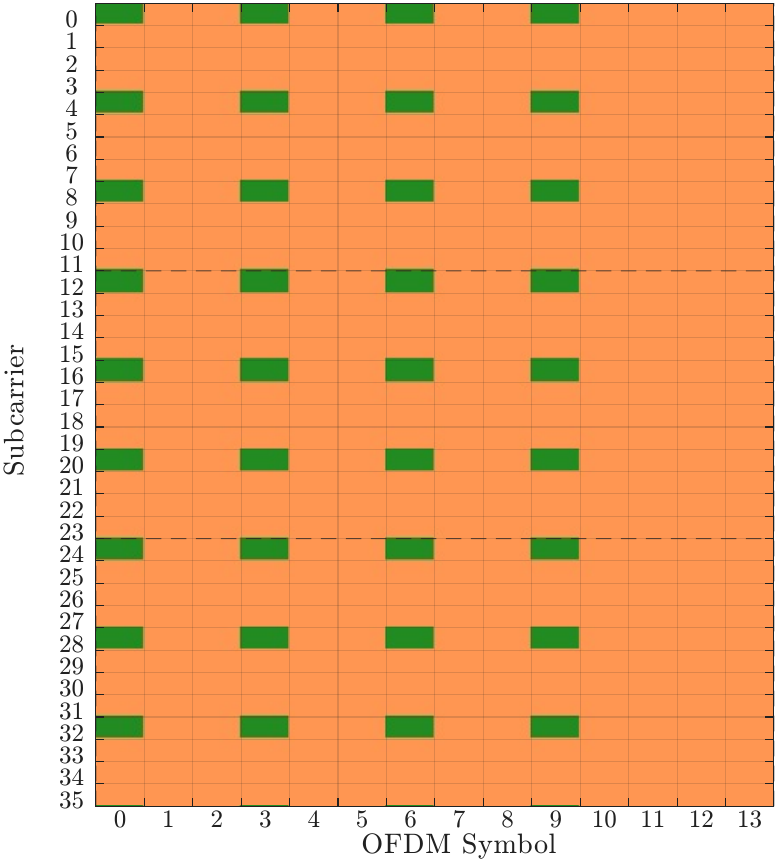}
  \caption{$\{ M_{\rm DDRS}, \KcombDDRS \} = \{4,4\}$.}
  \label{fig:GRID_Comb4_1Slot}  
  
\end{subfigure}
\hfill
\begin{subfigure}[t]{0.24\textwidth}
  \centering
  \includegraphics[width=\linewidth]{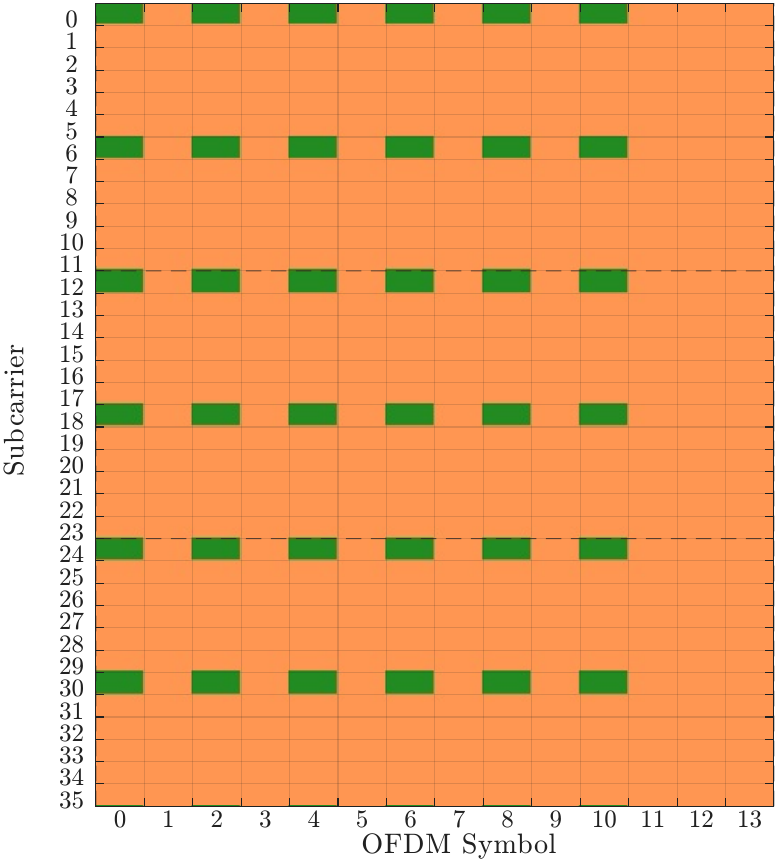}
  \caption{$\{ M_{\rm DDRS}, \KcombDDRS \} =  \{6,6\}$.\vspace{0.03cm}}
  \label{fig:GRID_Comb6_1Slot}  
  
\end{subfigure}
\hfill
\begin{subfigure}[t]{0.24\textwidth}
  \centering
  \includegraphics[width=\linewidth]{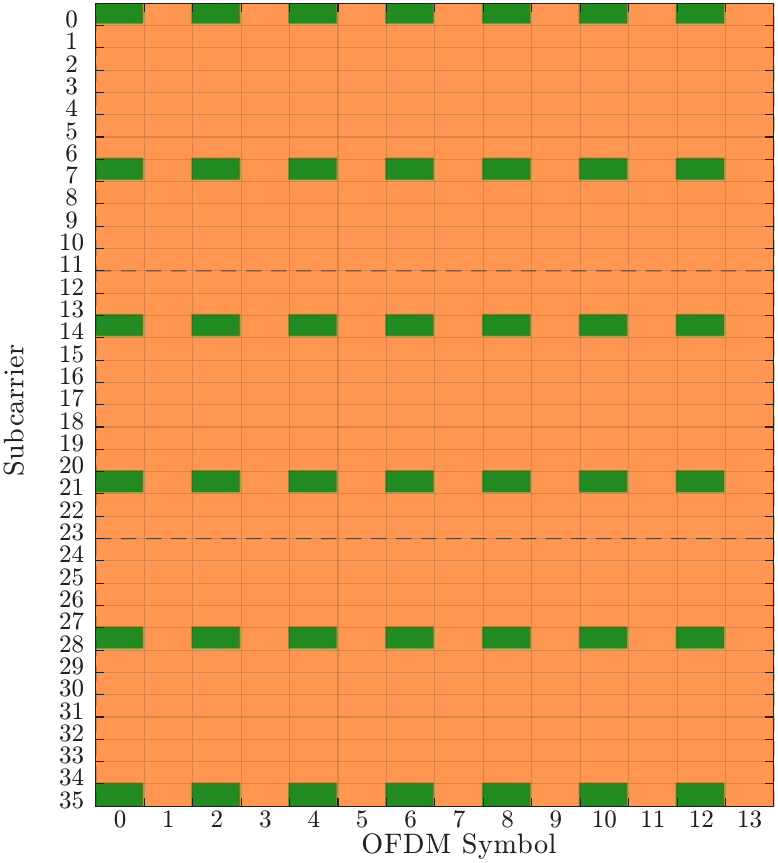}
  \caption{$\{ M_{\rm DDRS}, \KcombDDRS \} =  \{7,7\}$.}
  \label{fig:GRID_Comb7_1Slot}  
  
\end{subfigure}
\hfill
\begin{subfigure}[t]{0.24\textwidth}
  \centering
  \includegraphics[width=\linewidth]{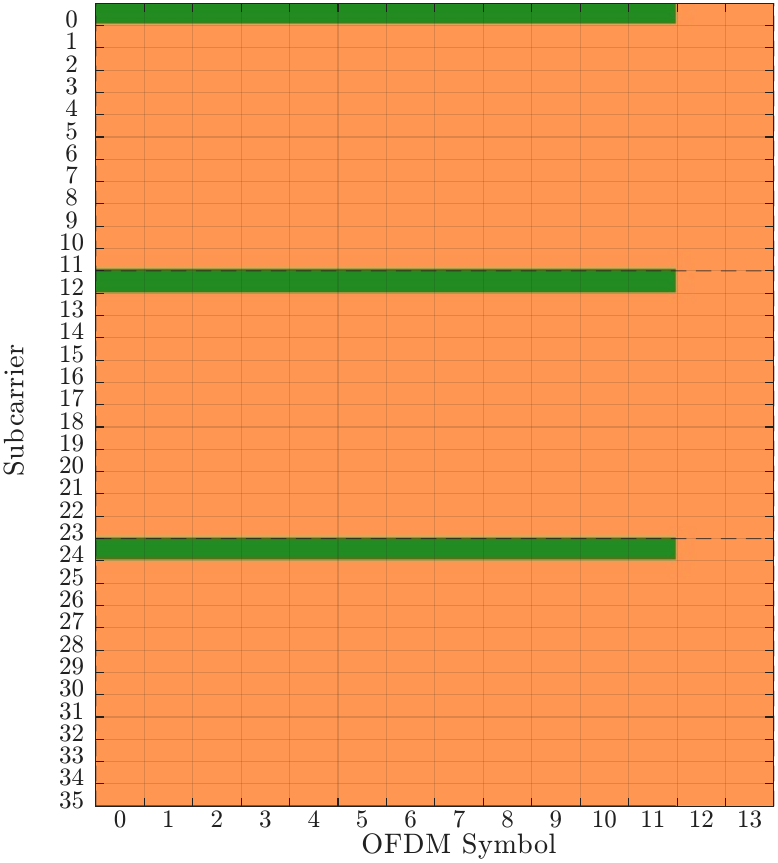}
  \caption{$\{ M_{\rm DDRS}, \KcombDDRS \} =  \{12,12\}$.}
  \label{fig:GRID_Comb12_1Slot}  
  
\end{subfigure}
\hfill
\begin{subfigure}[t]{0.24\textwidth}
  \centering
  \includegraphics[width=\linewidth]{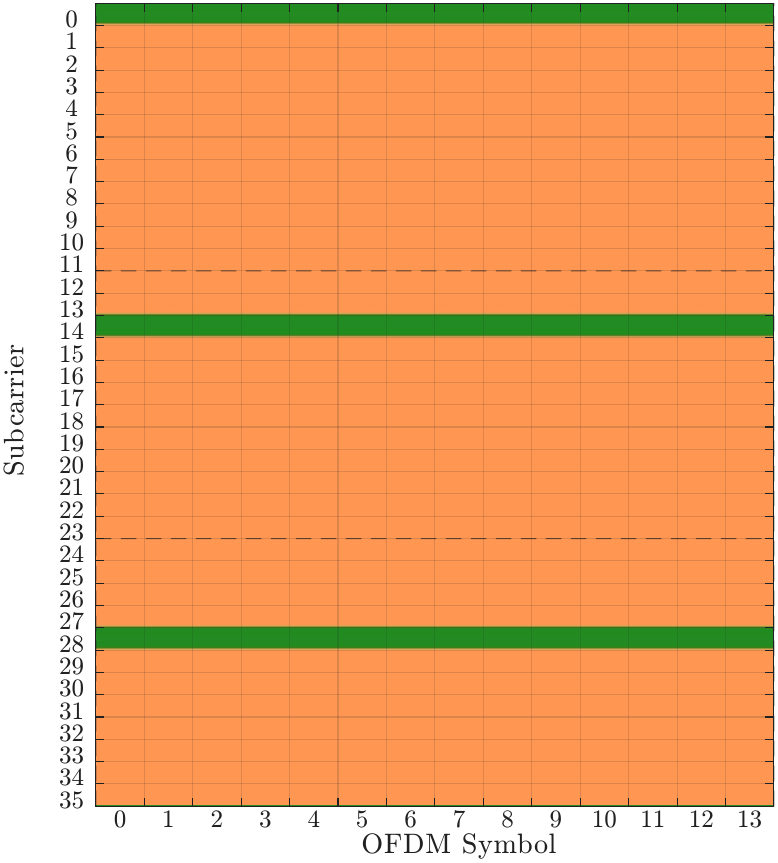}
  \caption{$\{ M_{\rm DDRS}, \KcombDDRS \} = \{14,14\}$.}
  \label{fig:GRID_Comb14_1Slot}  
  
\end{subfigure}
\caption{Patterns of the proposed \ac{DDRS} signal for sensing purposes.}
\label{fig:DDRS_pattern}
\end{figure}

Fig. \ref{fig:DDRS_pattern}  shows the defined \ac{DDRS} patterns. All configurations have the same overhead as the ones of the \acp{PRS} in Fig. \ref{fig:GRID_Comb_2_4_6_12_1Slot_PRS}: 7.14\%.

It is worth noting that the configurations $\left\{12,12\right\}$ and $\left\{14,14\right\}$ have the same symbol separation, $\Mcomb^{\mathrm{DDRS}}=1$, but the latter occupies all the symbols within a slot. The same applies to the configurations $\left\{6,6\right\}$ and $\left\{7,7\right\}$, where $\Mcomb^{\mathrm{DDRS}}=2$, although the latter does not cover all the symbols within the slot.

\section{Proposed range and velocity ML estimators}
\label{Sec:ML_estimators}
In this section we derive range and speed estimators for the signal in (\ref{eq:SignalModel:Y_qm_post_FFT}) and the full-slot sensing pattern, hence, $\Kmset=\mathcal{K}$. Since the range and velocity information is embedded in the phase of the received constellation symbol, let us define
\begin{equation}
    Z_{q,m} = Y_{q,m}^* X_{q,m} = \sqrt{\alpha_{\textrm{T}}} E_X  e^{j 2\pi \varphi_{q,m}(\tau_d, \vrad)} + W'_{q,m}, 
    \label{eq:MLestimators:Z_qm_definition}
\end{equation}
with $\varphi_{q,m}(\tau_d, \vrad)$ as given in (\ref{eq:SignalModel:Phase_of_Y_qm}) and $W'_{q,m} = X_{q,m} W_{q,m}^*$.

In order to estimate the radial-velocity, it can be observed that $\varphi_{q,m}(\tau_d, \vrad)$ can be expressed as
\begin{equation}
    \varphi_{q,m}(\tau_d, \vrad) =  f_{\vrad} m + \beta_q, 
    \label{eq:MLestimators:varphi_as_fq_kappa_q}
\end{equation}
where
\begin{equation}
    \begin{aligned}
        f_{\vrad} &= (\fc + \Delta fq) \frac{2\vrad}{c_0} L \Ts.\\
        \beta_q &= \Delta f q (\taud - \taus)  \\ 
        & + \left(\fc + \Delta f q\right) \frac{2\vrad}{c_0} \left(\ns + \Ncp + \frac{N-1}{2}\right)\Ts + \psi.
    \end{aligned}
    \label{eq:MLestimators:fq_kappa_q_definitions}
\end{equation}

Since $\vrad$ can be positive or negative, the span of normalized frequencies that can be unambiguously determined is $f_{\vrad} \in \left[ -1/2, 1/2\right)$. Hence, assuming that $\fc \gg \Delta f N_{\rm A}/2$, the range of velocities that can be estimated unambiguously is $\vrad \in [-\vrad^{\rm max},\vrad^{\rm max})$, with
\begin{equation}
\vrad^{\rm max}= \frac{c_0 N \Delta f}{4\fc L}.
    \label{eq:MLestimators:unambiguous_velocity_range}
\end{equation}

The \ac{ML} estimator of $f_{\vrad}$ from $Z_{q,m}$, assuming that both $\alpha_{\textrm{T}}$ and $\beta_q$ are unknown, can be obtained as in \cite[Eq.~(7.66)]{Kay1993}. These $N_{\rm A}$ estimates, one for each $q\in[-N_{\rm A}/2,N_{\rm A}/2-1]$, can then be averaged to obtain a single estimate of $\vrad$. However, the estimates obtained with this procedure suffer from the well-known threshold effect \cite{Steinhardt1985}, which severely limits the range of \ac{SNR} values over which it attains the \ac{CRLB}. To extend this range, the noise has to be reduced before the estimation. 

The following two-step iterative process can be employed to achieve this end. First, a coarse estimate of the radial-velocity, $\hat{v}_{\phi}^{\rm c}$, is obtained from the frequency-averaged value of $Z_{q,m}$ as
\begin{equation}
\begin{aligned}
 \overline{Z}_m^{\rm c} = \frac{1}{N_{\rm A}} \sum_{q=-N_{\rm A}/2}^{N_{\rm A}/2-1}\!\!\!\!\! Z_{q,m} 
        =& \sqrt{\alpha_{\textrm{T}}}E_X \underbrace{\frac{1}{N_{\rm A}}\frac{\sin(\pi\gamma_{\tau_d}N_{\rm A})}{\sin(\pi\gamma_{\tau_d})}}_{D_{N_{\rm A}}(2\pi\gamma_{\tau_d})}\\
        &\times e^{j2\pi\left(F_{\vrad}^{\rm c}m+\eta^{\rm c}\right)} + \overline{W}_m^{\rm c},
\end{aligned}
\label{eq:Z_averaged}
\end{equation}
where $\overline{W}_m^{\rm c} \sim CN(0,\sigma_w^2/N_{\rm A})$, $D_{N_{\rm A}}(x)$ denotes the Dirichlet kernel with argument
\begin{equation}
\gamma_{\tau_d} = \Delta f\left( \tau_d - \tau_{\textrm{R}} + \frac{2v_{\phi}}{c_0}\delta_mT_s \right),
\end{equation}
and where the frequency $F_{\vrad}^{\rm c}$ and $\eta^{\rm c}$ are given by
\begin{equation}
\begin{aligned}
    F_{\vrad}^{\rm c} &=  \frac{2v_{\phi}}{c_0} L T_s \left( f_c  -  \frac{\Delta f}{2}    \right),\\
    \eta^{\rm c}&= \left( f_c - \frac{\Delta f}{2}    \right) \times \left( \tau_d + \frac{2v_{\phi}}{c_0}\left(\ns + \Ncp + \frac{N-1}{2}\right)T_s \right)\\
    &+ \Delta f  \tau_{\textrm{R}}/2  - \frac{\phi_{\rm S}}{2\pi}.
\end{aligned}
\end{equation}

A coarse estimate of the radial-velocity, $\hat{v}_{\phi}^{\rm c}$, is then obtained by firstly computing the \ac{ML} estimate of $F_{\vrad}$ ($\eta^{\rm c}$ is unknown) as \cite[Eq.~(7.66)]{Kay1993}
\begin{equation}
    \hat{F}_{\vrad}^{\rm c} = \operatorname*{argmax}_{f} \left\{ \left| \sum_{m=0}^{M-1} \overline{Z}_m^{\rm c} e^{-j2\pi f m} \right|^2 \right\},
    \label{eq:MLestimators:f_q_estimator_periodogram}
\end{equation}
which can be implemented by means of a \ac{DFT}-based periodogram and a subsequent fine search using a numerical method \cite{Rife1974}, and then computing
\begin{equation}
    \hat{v}_{\phi}^{\rm c}=  \frac{\hat{F}_{\vrad}^{\rm c} c_0}{2 (f_c - \Delta f/2) L T_s}.
    \label{eq:MLestimators:velocity_estimator}
\end{equation}

Since the noise variance in $\overline{Z}_m^{\rm c}$ is $N_{\rm A}$ times lower than in $Z_{q,m}$, the threshold effect is displaced to lower \ac{SNR} values by $10\log_{10}(N_{\rm A})$ (dB). However, the amplitude of $\overline{Z}_m^{\rm c}$ depends on the range of the \ac{ST} by means of the Dirichlet kernel $D_{N_{\rm A}}(2\pi\gamma_{\tau_d})$. This makes the actual \ac{SNR} in (\ref{eq:Z_averaged}) to vary periodically with the range of the \ac{ST}. This problem will be corrected in the second phase of the algorithm. 

Next, the coarse estimate of the range, $\hat{d}^{\rm c}$ is obtained. To this end, the terms associated to the radial-velocity in $Z_{q,m}$ are eliminated as 
\begin{equation}
H_{q,m}^{\rm c}=Z_{q,m}e^{-j2\pi(\fc + \Delta f  q)\frac{2 \hat{v}_{\phi}}{c_0}\delta_m\Ts} + W^{\prime\prime}_{q,m},
\label{eq:H_qm}
\end{equation}
where $ W^{\prime\prime}_{q,m} = X_{q,m} W_{q,m}^*e^{-j2\pi(\fc + \Delta f  q)\frac{2 \hat{v}_{\phi}}{c_0}\delta_m\Ts}$.
In order to mitigate the threshold effect in the range estimation, a time averaging of $H_{q,m}^{\rm c}$ is performed. Assuming perfect estimation of the radial-velocity, this yields
\begin{equation}
\overline{H}_{q}^{\rm c}=\frac{1}{M} \sum_{m=0}^{M-1} H_{q,m}^{\rm c}=\sqrt{\alpha_{\textrm{T}}}E_X e^{j2\pi\psi}e^{j2\pi\overbrace{\scriptstyle \Delta f \left( \taud -\taus \right)}^{F_d^{\rm c}}q}+ \overline{W}_q^{\rm c},
\label{eq:H_q}
\end{equation}
where  $\overline{W}_q^{\rm c}\sim CN(0,\sigma_w^2/M)$.

The \ac{ML} estimate of $F_d^{\rm c}$ is then obtained from the observations $H_q^{\rm c}$ as
\begin{equation}
    \hat{F}_d^{\rm c} = \operatorname*{argmax}_{f} \left\{ \left| \sum_{q=0}^{N_{\rm A}-1} H_q^{\rm c} e^{-j2\pi f q} \right|^2 \right\},
    \label{eq:MLestimators:f_d_estimator_periodogram}
\end{equation}
and the coarse range estimate as 
\begin{equation}
    \hat{d}^{c} = \frac{\hat{F}_d^{\rm c}}{\Delta f } + \taus.
    \label{eq:d_estimator}
\end{equation}

The range of the \ac{ST}, $d$, is a positive value, so the span of normalized frequency, $f$, where the maximum of the periodogram in (\ref{eq:MLestimators:f_d_estimator_periodogram}) can be determined without ambiguity is $f \in \left[0, 1\right)$. In consequence, the maximum range that can be unambiguously estimated is given by
\begin{equation}
d_{\textrm{max}} = \frac{c_0}{2}\left( \frac{1}{\Delta f} + \taus \right).
\label{eq:d_max}
\end{equation}

Now, in order to obtain a refined estimate of the radial-velocity, the range dependence in $Z_{q,m}$ can be eliminated, which removes the Dirichlet kernel that appeared in the averaging of $Z_{q,m}$ performed to mitigate the threshold effect. To this end, 
\begin{equation}
\begin{aligned}
 \overline{Z}_m^{\rm r} &= \frac{1}{N_{\rm A}} \sum_{q=-N_{\rm A}/2}^{N_{\rm A}/2-1}\!\!\!\!\! Z_{q,m} e^{-j2\pi\Delta f q(2\hat{d}^{\rm c}/c_0-\taus)} \\
 &=\sqrt{\alpha_{\textrm{T}}}E_X e^{j2\pi\left(F_{\vrad}^{\rm r}m+\eta^{\rm r}\right)} + \overline{W}_m^{\rm r},
\end{aligned}
\label{eq:Zr_averaged}
\end{equation}
The \ac{ML} estimate of $F_{\vrad}^{\rm r}$ is now computed analogously to (\ref{eq:MLestimators:f_q_estimator_periodogram}), and a refined version of the radial-velocity, $\hat{v}_{\phi}^{\rm r}$, is obtained as in (\ref{eq:MLestimators:velocity_estimator}). 

Finally, a refined version of the range estimate, $\hat{d}^{\rm r}$, is obtained. To this end, the effect of $\vrad$ in $Z_{q,m}$ is firstly compensated as in (\ref{eq:H_qm}), yielding $H_{q,m}^{\rm r}$, whose time-averaged value, $\overline{H}_{q}^{\rm r}$, is obtained as in (\ref{eq:H_q}). Finally, $\hat{d}^{\rm r}$ is obtained proceeding analogously to (\ref{eq:MLestimators:f_d_estimator_periodogram}) and (\ref{eq:d_estimator}). 

\begin{algorithm}
\caption{Two-step iterative range and speed estimation algorithm}
  \begin{algorithmic}[1]
    \Statex \hspace*{-\algorithmicindent}\textbf{First iteration}
    \State Compute $Z_{q,m}$ from $Y_{q,m}$ as in (\ref{eq:MLestimators:Z_qm_definition})
    \State Calculate $\overline{Z}_m^{\rm c}$ by averaging $Z_{q,m}$ over the active subcarriers as in (\ref{eq:Z_averaged})
    \State Obtain a coarse estimate of the radial-velocity, $\hat{v}_{\phi}^{\rm c}$, from $\overline{Z}_m^{\rm c}$ using (\ref{eq:MLestimators:f_q_estimator_periodogram}) and (\ref{eq:MLestimators:velocity_estimator})
    \State Eliminate the velocity terms in $\varphi_{q,m}(\tau_d, \vrad)$ by computing $H_{q,m}^{\rm c}$ as in (\ref{eq:H_qm})
    \State Obtain a coarse estimate of the range, $\hat{d}^{\rm c}$, from $\overline{H}_q^{\rm c}$ in (\ref{eq:H_q}) using 
    (\ref{eq:MLestimators:f_d_estimator_periodogram}) and (\ref{eq:d_estimator})    
    \Statex \hspace*{-\algorithmicindent}\textbf{Second iteration}
    \State Eliminate the range terms in $Z_{q,m}$ and compute its time-averaged value, $\overline{Z}_m^{\rm r}$, as in (\ref{eq:Zr_averaged})
    \State Obtain a refined estimate of the radial-velocity, $\hat{v}_{\phi}^{\rm r}$, from $\overline{Z}_m^{\rm r}$ analogously to (\ref{eq:MLestimators:f_q_estimator_periodogram}) and (\ref{eq:MLestimators:velocity_estimator})
    \State Eliminate the velocity terms in $\varphi_{q,m}(\tau_d, \vrad)$ by computing $H_{q,m}^{\rm r}$ and obtaining its frequency-averaged value, $\overline{H}_{q}^{\rm r}$, analogously to (\ref{eq:H_q})
    \State Obtain a refined estimate of the range, $\hat{d}^{\rm }$, from $\overline{H}_q^{\rm r}$ proceeding as in 
    (\ref{eq:MLestimators:f_d_estimator_periodogram}) and (\ref{eq:d_estimator})    
  \end{algorithmic} 
  \label{alg:two_step_estimation}
\end{algorithm}

As mentioned, the employed \ac{ML} estimators in (\ref{eq:MLestimators:f_q_estimator_periodogram}) and (\ref{eq:MLestimators:f_d_estimator_periodogram}) can be implemented by means of a \ac{DFT} and a subsequent refinement using a numerical method. The size of the \ac{DFT}, denoted as $\Nper$, determines the resolution of the estimates, as the refinement is carried out around a normalized frequency constrained to a discrete set. Since the span of $F_{\vrad} \in [-1/2,1/2)$ and $F_{d} \in [0,1)$ is equally spaced divided into $\Nper$ points, i.e., $\Delta F = 1 / \Nper$,  the following velocity and range resolutions are obtained under the assumption that $\fc \gg \Delta f N_{\rm A}/2$
\begin{equation}
    \begin{aligned}
        &\Delta \vrad = \frac{c_0 N \Delta f}{2\fc\Nper L}\\
        &\Delta d = \frac{c_0}{2\Nper \Delta f }
    \end{aligned}.
    \label{eq:MLestimators:resoluciones}
\end{equation}

\section{Numerical results and discussion}
\label{Sec:NumericalResults&Discussion}
This section shows the highest accuracy that can be achieved in the range and radial-speed estimation of an \ac{UAV} using monostatic sensing. Next, the suitability of the \ac{PRS} for sensing purposes is evaluated and compared against the performance of the proposed \ac{DDRS} pattern, to assess whether the introduction of a new sensing-oriented reference signal is justified, or whether the improvements reported for other seemingly better patterns are not practically significant. Finally, since the aforementioned results are obtained by means of the \ac{CRLB}, but there is no guarantee that an efficient estimator can be obtained for the considered problem, we show that the performance given by the two-step iterative estimator proposed in Section \ref{Sec:ML_estimators} is efficient and compare it to the performance of a plain \ac{ML} estimator.

In all cases, the accuracy of the range and radial-velocity is obtained for estimators that use sensing resources from a number of slots ranging from 1 to 20. Increasing the number of slots used in the estimation improves performance at the cost of an increased complexity, in particular the memory required to store the complex values $Y_{q,m}$ to be used by the estimator. 

\subsection{System parameters}
This work uses the system parameters defined in \cite{3gpp_R1_2507427_2025} for the considered use case and that are summarized in Table \ref{table:paramaters_UAV}. 

\begin{table}[ht]
\caption{System parameters}
    \centering
    \begin{tabular}{c c c}
    \cline{1-3}
      \textbf{Parameter} & \textbf{Description} & \textbf{Value}\\ 
      \cline{1-3}
      $\mu$ & Numerology index & 1 \\
      $f_c$ & Carrier frequency & 4 GHz\\
      $\Delta f$ & \ac{SCS} &30 kHz \\
      BW & Channel bandwidth &100 MHz\\
      $N$ & Number of carriers & 4096 \\ 
      $N_{\rm cp}$ & Cyclic prefix (samples)& 288 samples\\
      $T_{\rm s}$ & Sampling period& 8.14 ns\\
      $N_{\rm RB}$ &  Number of active \acp{RB}& 273\\
      $N_{\rm A}$ &  Number of active subcarriers& 3276\\
      \cline{1-3}
    \end{tabular}
    \label{table:paramaters_UAV}
\end{table}

The employed \ac{RCS} model is based on \cite[Sec.~7.9.2.1 (model~1)]{3gpp_tr_38_901_v19_1_0}, which states that it can be expressed as the product of three terms, $\sigma_{\textrm{RCS}} = \sigma_{\textrm{M}} \sigma_{\textrm{D}} \sigma_{\textrm{S}}$, where ${\sigma_{\textrm{M}} = -12.81}$ dBsm is deterministic, $\sigma_{\textrm{D}}=1$ when it is independent of the angle of incidence, and $\sigma_{\textrm{S}}$ is a log-normal random variable such that $10\textrm{log}(\sigma_{\textrm{S}})~\sim~\mathcal{N}(\mu_{\sigma_{\textrm{S\_dB}}}, \sigma_{\sigma_{\textrm{S\_dB}}}^2)$. In this work, the latter is fixed to its mean value, i.e., $\sigma_{\mathrm{S}}(\mathrm{dB}) = \mu_{\sigma_{\textrm{S\_dB}}}$, whose expression is related to its variance through
\begin{equation}
        \mu_{\sigma_{\textrm{S\_dB}}} = -\frac{\textrm{ln}(10)}{20} \sigma_{\sigma_{\textrm{S\_dB}}}^2,
    \label{eq:muSigma_S}   
\end{equation}
with $\sigma_{\sigma_{\textrm{S\_dB}}} = 3.74$ dB for small \acp{UAV}.

The inter-BS distance of 500 m agreed in \cite{3gpp_R1_2507427_2025} is assumed, which leads to a maximum coverage distance of roughly 290 m. Since the \ac{ST} can be up to 300 m above the \ac{BS}, the maximum range is just below 420 m. Hence, the accuracy for \acp{ST} located at a maximum distance of 440 m is evaluated. According to \cite{3gpp_R1_2506479_2025}, a maximum speed of 50 m/s (180 km/h) is considered. 

The \acp{KPI} defined in \cite{3gpp_R1_2509243_2025} for the considered use case are given in Table \ref{table:PerfObjectivesISAC}. Regarding range accuracy, since the distance $d$ considered in this work includes both horizontal and vertical components, a 10 m accuracy requirement is imposed on $d$ to ensure compliance in all cases. Interestingly, the values in Table \ref{table:PerfObjectivesISAC} limit the maximum number of slots that can be used in the estimation. For instance, in 400 slots (0.2 s), a \ac{ST} moving at 50 m/s the \ac{ST} has displaced 10 m (the accuracy limit) from its position. Fortunately, a lower number of slots is actually required to meet the requirements, as it will be shown. 

\begin{table}[!b]
\caption{\acp{KPI} adopted for evaluation purpose of \ac{NR} \ac{ISAC}.}
    \label{table:PerfObjectivesISAC}
    \centering
    \begin{tabular}{c c}
    \toprule
        \textbf{\ac{KPI}} &  \textbf{Required accuracy and confidence level} \\
        \midrule                
        Horizontal Positioning	& 10 m with confidence level 90 \% \\
        \hline        
        Vertical Positioning & 10 m with confidence level 90 \% \\
        \hline        
        Velocity	& 5 m/s with confidence level 90 \% \\
        \bottomrule
    \end{tabular}    
\end{table}
%

\subsection{Sensing resolution and maximum values}

Expressions (\ref{eq:MLestimators:unambiguous_velocity_range}) and (\ref{eq:d_max}) gave the maximum radial-velocity and range values that can be unambiguously estimated, while expression (\ref{eq:MLestimators:resoluciones}) provided the resolution that can be achieved when estimating $d$ and $\vrad$ using an $\Nper$-\ac{DFT}. For illustrative purposes, Table \ref{table:maximum_unamb_estim} reports the maximum unambiguous values for range and velocity obtained using the system parameters in Table \ref{table:paramaters_UAV}. Similarly, Table \ref{table:estimate_resolution_Nper} provides the attainable resolution for different values of $\Nper$. Note that the indicated limit for range estimation corresponds to the scenario where $\taus=0$, i.e., frame synchronization is not performed. Otherwise, larger ranges can, in principle, be estimated without ambiguity. However, in practice the maximum range is constrained by the sensitivity of the receiver.

\begin{table}[!b]
\caption{Maximum unambiguous values of the range and radial-velocity estimates for the system parameters in Table \ref{table:paramaters_UAV}. }
    \label{table:maximum_unamb_estim}
    \centering
    \begin{tabular}{c c}
    \toprule
        \textbf{Magnitude} &  \textbf{Maximum value} \\
        \midrule                
        Range ($d_{\mathrm{max}}$)	&  4996.5 m \\
        \hline        
        Velocity ($v_{\phi}^{\mathrm{max}}$) &  525.2 m/s \\
        \bottomrule
    \end{tabular}    
\end{table}

\begin{table}[!b]
\caption{Resolution values of the range and radial-velocity estimates when using \ac{DFT}-based estimators with $N_{\textrm{per}}$ samples and the system parameters in Table \ref{table:paramaters_UAV}.}
    \label{table:estimate_resolution_Nper}
    \centering
    \begin{tabular}{c c c c c c}
    \toprule
        $N_{\textrm{per}}$ & 16 & 256 & 4096 & 65536 \\
        \midrule                
        $\Delta d$ (m)	& 313.3 & 19.52 & 1.220 & 0.076\\
        \hline        
         $\Delta v_{\phi}$ (m/s) & 65.6 & 4.10 & 0.256 & 0.016\\
        \bottomrule
    \end{tabular}    
\end{table}

\subsection{Accuracy assessment}
\begin{figure}[ht!]
\centering
\begin{subfigure}[t]{0.45\textwidth}
  \centering
  \includegraphics[width=\linewidth]{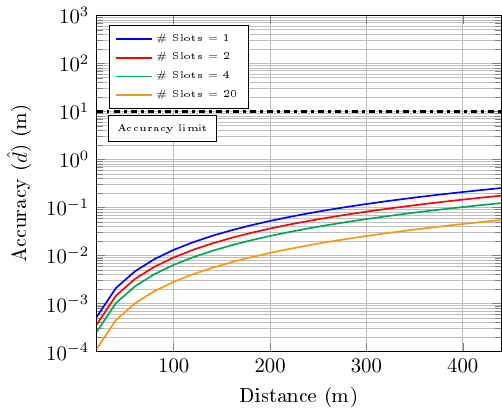}
   \caption{Highest accuracy in range estimation.}
  \label{fig:Comparison_CRLB_d_1_20_Slots_FULL}   
\end{subfigure}
\hfill
\begin{subfigure}[t]{0.45\textwidth}
  \centering
  \includegraphics[width=\linewidth]{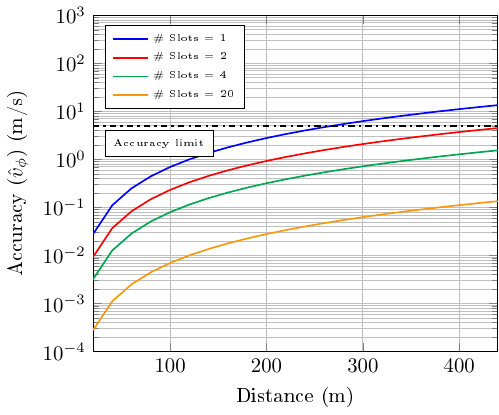}
  \caption{Highest accuracy in radial-velocity estimation.}
  \label{fig:Comparison_CRLB_v_1_20_Slots_FULL}   
\end{subfigure}
\caption{Maximum achievable accuracy in range and radial-velocity estimations using the full-slot sensing pattern.}
\label{fig:Comparison_CRLB_v_d_1_20_Slots_FULL}
\end{figure}

First, accuracy values given by (\ref{eq:CRLB:Accuracy_expression}) using the variance given by the \ac{CRLB} in (\ref{eq:CRLB:CRLB_full_slot_compact}) for the full-slot pattern and the confidence levels in Table \ref{table:PerfObjectivesISAC} are presented. Fig. \ref{fig:Comparison_CRLB_v_d_1_20_Slots_FULL} shows the results obtained when the estimation is computed over 1, 2, 4, and 20 time slots. As seen, the distance accuracy requirement is easily satisfied with a single time slot, whereas at least two time slots are necessary to meet the velocity accuracy requirement.

\begin{figure}[b!]
\centering
\begin{subfigure}[t]{0.45\textwidth}
  \centering
  \includegraphics[width=\linewidth]{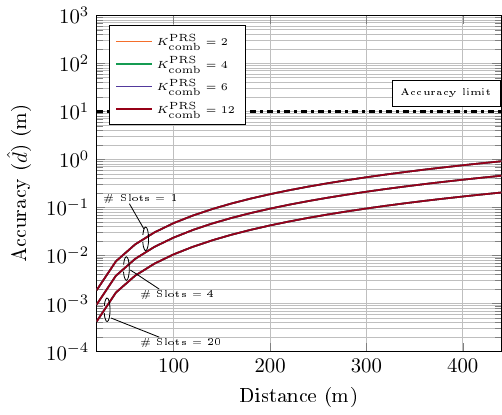}
  \caption{Highest accuracy in range estimation.}
  \label{fig:Comparison_CRLB_d_1_4_20_Slots_PRS}   
\end{subfigure}
\hfill
\begin{subfigure}[t]{0.45\textwidth}
  \centering
  \includegraphics[width=\linewidth]{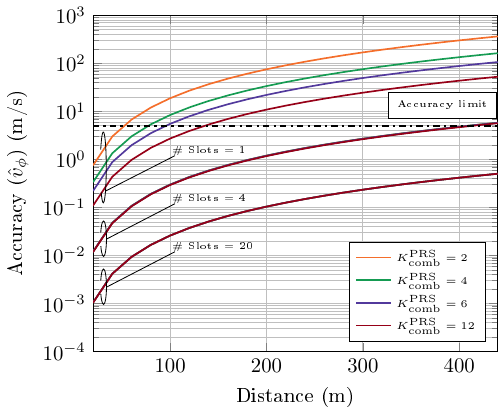}
  \caption{Highest accuracy in radial-velocity estimation.}
  \label{fig:Comparison_CRLB_v_1_4_20_Slots_PRS} 
\end{subfigure}

\caption{Maximum achievable accuracy in range and radial-velocity estimations using different \ac{PRS} sensing patterns.}
\label{fig:Comparison_CRLB_v_d_1_4_20_Slots_PRS}
\end{figure}

As expected, lower performance is obtained when using \acp{PRS}. Fig. \ref{fig:Comparison_CRLB_v_d_1_4_20_Slots_PRS} shows the accuracy bounds (derived from the \ac{CRLB}) when the range and radial-velocity are estimated over the \ac{PRS} resources from $1, \, 4$ and $20$ time slots. It can be observed that a single time slot is more than sufficient to meet the distance accuracy requirement, however, more than 4 time slots are now required to meet the velocity accuracy requirements. In practice, the performance is independent of $\KcombPRS$ when multiple time slots are employed. However, the accuracy strongly depends on the \ac{PRS} pattern when a single time slot is used: the larger $\KcombPRS$, the better the accuracy. The rationale for this behavior is that the radial-velocity information induces a phase shift in the received symbols, $Y_{q,m}$, which primarily varies with the symbol index $m$, as shown in (\ref{eq:MLestimators:varphi_as_fq_kappa_q}), and the time extent of the \ac{PRS} pattern increases with $\KcombPRS$, as illustrates in Fig. \ref{fig:GRID_Comb_2_4_6_12_1Slot_PRS}. This can also be explained in terms of the \ac{CRLB} in (\ref{eq:CRLB:CRLB_full_slot_compact}), which shows that the performance improves with increasing variance of the indices of the \ac{OFDM} symbols used in the estimation, and the latter increases with $\KcombPRS$.

\begin{figure}[b!]
\centering
\begin{subfigure}[t]{0.45\textwidth}
  \centering
  \includegraphics[width=\linewidth]{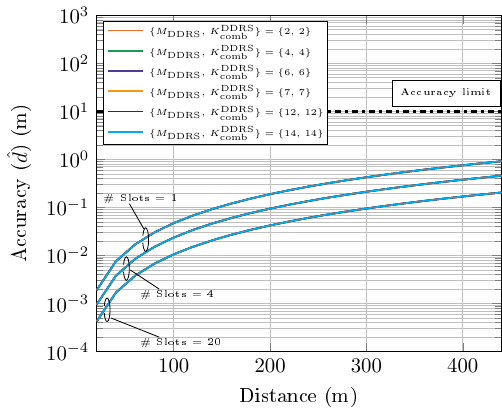}
  \caption{Highest accuracy in range estimation.}
  \label{fig:Comparison_CRLB_d_1_4_20_Slots_DDRS}  
\end{subfigure}
\hfill
\begin{subfigure}[t]{0.45\textwidth}
  \centering
  \includegraphics[width=\linewidth]{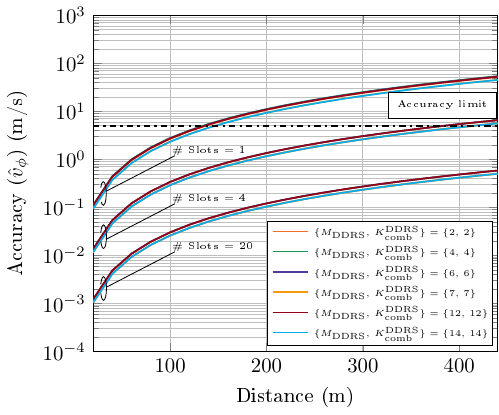}
  \caption{Highest accuracy in radial-velocity estimation.}
  \label{fig:Comparison_CRLB_v_1_4_20_Slots_DDRS} 
\end{subfigure}
\caption{Maximum achievable accuracy in range and radial-velocity estimations using different \ac{DDRS} sensing patterns.}
\label{fig:Comparison_CRLB_v_d_1_4_20_Slots_DDRS}
\end{figure}

Now, the performance obtained with the proposed \ac{DDRS} patterns is assessed. Fig. \ref{fig:Comparison_CRLB_v_d_1_4_20_Slots_DDRS} depicts the accuracy bounds for the range and radial-velocity obtained when the \ac{DDRS} resources in $1, \, 4$ and $20$ time slots are used. As observed, similar accuracy levels are obtained for all \ac{DDRS} patterns. The configurations $\{ M_{\rm DDRS}, \KcombDDRS \} = \{\{2,2\}, \{7,7\}, \{14,14\}\}$ provide slightly better radial-velocity accuracy when $4$ and $20$ time slots are considered. In contrast, for a single time slot, the configurations $\{\{7,7\}, \{14,14\}\}$ achieve the best performance. Interestingly, obtained values are similar to the ones given by the \acp{PRS} when the estimation is performed using multiple time slots. When a single time slot is used, the range estimation accuracy is similar to that obtained with the \acp{PRS}, whereas the radial-velocity accuracy becomes almost independent of the \ac{DDRS} pattern and matches the best-case performance of the \ac{PRS} pattern obtained with $K^{\mathrm{PRS}}_{\mathrm{comb}} = 12$. This result is in agreement with the \ac{CRLB} in (\ref{eq:CRLB:CRLB_full_slot_compact}), since the variance of the symbol indices of the \ac{PRS} is very small when a single time slot is employed, but similar to that of the \ac{DDRS} when multiple slots are used.

Since the accuracy values obtained from a single time slot are insufficient to meet the radial-velocity accuracy requirement in Table \ref{table:PerfObjectivesISAC}, practical estimators have to resort to multiple slots. Hence, presented results suggest that defining a new reference signal for sensing purposes may be unnecessary when the \ac{5G} frame structure is employed. 

\begin{figure}[b!]
\centering
\begin{subfigure}[t]{0.45\textwidth}
  \centering
  \includegraphics[width=\linewidth]{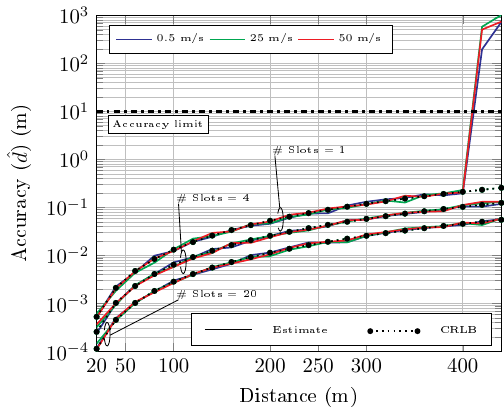}
  \caption{Accuracy obtained when estimating the range using the proposed two-step iterative algorithm.}
  \label{fig:PatchedChannel_Distance_severalVelocities_Iterative_1_4_20_Slots}  
\end{subfigure}
\hfill
\begin{subfigure}[t]{0.45\textwidth}
  \centering
  \includegraphics[width=\linewidth]{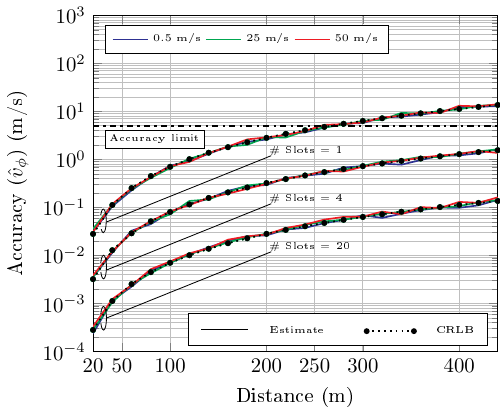}
\caption{Accuracy obtained when estimating the radial-velocity using the proposed two-step iterative algorithm.}  \label{fig:PatchedChannel_Velocity_severalVelocities_Iterative_1_4_20_Slots}   
\end{subfigure}
\caption{Accuracy obtained when estimating the radial-velocity using the proposed two-step iterative algorithm and the full-slot sensing pattern.}
\label{fig:PatchedChannel_Vel_Dist_severalVelocities_Iterative_1_4_20_Slots}
\end{figure}

Accuracy values presented so far have been derived from the variance given by the \ac{CRLB}. We now prove that the estimator proposed in Algorithm \ref{alg:two_step_estimation} attains this bound over a wide range of distances. To this end, Fig.~\ref{fig:PatchedChannel_Vel_Dist_severalVelocities_Iterative_1_4_20_Slots} depicts the accuracy of the range and radial-velocity estimations obtained when the full-slot sensing pattern is employed. Presented values are computed from (\ref{eq:CRLB:Accuracy_expression}) using the bias and variance of the estimates obtained by means of Monte Carlo simulations. Results are shown for \acp{ST} with different radial velocities and three observation sets ($1$, $4$, and $20$ time slots). As seen, the proposed estimator is efficient over the required range of distances (up to $420$ m), except for the estimates obtained using a single time slot, for which the threshold effect limits the range over which the accuracy requirement is met to a maximum of $400$ m. 

\begin{figure}[b!]
\centering
\begin{subfigure}[t]{0.45\textwidth}
  \centering
  \includegraphics[width=\linewidth]{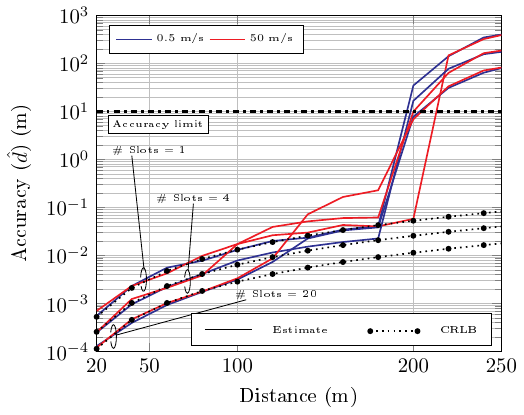}
  \caption{Accuracy obtained when estimating the range using a plain \ac{ML} estimator.} 
  \label{fig:Accuracy_D_vs_Distance_FullSlot_Slots_1_2_4_20_1_kmh}  
\end{subfigure}
\hfill
\begin{subfigure}[t]{0.45\textwidth}
  \centering
  \includegraphics[width=\linewidth]{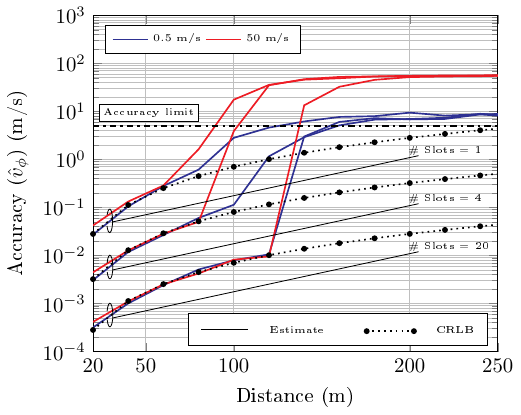}
  \caption{Accuracy obtained when estimating the radial-velocity using a plain \ac{ML} estimator.} 
  \label{fig:Accuracy_Vel_vs_Distance_FullSlot_Slots_1_2_4_20_1_kmh} 
\end{subfigure}
\caption{Accuracy of the range and radial-velocity estimation obtained with plain \ac{ML} estimators and the full-slot sensing pattern.}
\label{fig:Accuracy_vel_D_vs_Distance_FullSlot_Slots_1_2_4_20_1_kmh}
\end{figure}

Finally, to highlight the improvement of the proposed two-step iterative estimator over plain \ac{ML} ones, Fig. \ref{fig:Accuracy_vel_D_vs_Distance_FullSlot_Slots_1_2_4_20_1_kmh} shows the accuracy achieved by the latter, as defined in the paragraph immediately following (\ref{eq:MLestimators:unambiguous_velocity_range}), when using the full-slot sensing pattern. It can be observed that the threshold effect appears at short distances, preventing the performance requirement from being met over the considered range.

\section{Conclusion}
\label{Sec:Conclusion}

This work has determined the theoretical accuracy limits for range and velocity estimation of a single \ac{UAV} using \ac{5G} signals within a monostatic sensing framework. The compact \ac{CRLB} expressions derived herein highlight the fundamental trade-offs between estimation accuracy and system parameters. These variance bounds are then translated into accuracy values and compared with the requirements specified by the \ac{3GPP}. The obtained results indicate that, when all the resources within a slot are used for sensing purposes, the estimator must exploit information from at least two slots to meet the accuracy limit for velocity estimation, whereas a single slot suffices to satisfy the accuracy requirement for range estimation.

The accuracy achieved using the \ac{5G} \ac{NR} \ac{PRS}, which is the most suitable standardized reference signal for sensing purposes, is also evaluated and compared with that obtained using a reference signal specifically designed for sensing. The presented results show that, when estimation is performed over the \ac{PRS} resources of a single slot, the most suitable \ac{PRS} pattern is the one with a comb size equal to 12, $K_{\mathrm{comb}}^{\mathrm{PRS}}=12$. However, negligible differences between patterns are observed when estimation is carried out over multiple slots. Furthermore, since it has also been shown that more than four slots are required to meet the radial-velocity accuracy requirements imposed by the \ac{3GPP}, in practice, this implies that the \ac{PRS} yields nearly the same accuracy as specifically designed signals, which outperform it in single-slot estimation.

Although, in principle, there is no guarantee that a practical estimator attaining the \ac{CRLB} can be designed, we propose a two-step iterative range and radial-velocity estimator that is efficient over a significantly wider range of distances than conventional maximum-likelihood (ML) estimators. The presented results show that the latter are strongly affected by the well-known threshold effect, which limits the distance range over which the accuracy requirements imposed by the \ac{3GPP} are satisfied to less than 200 m for range estimation and less than 130 m for radial-velocity estimation.

\bibliographystyle{ieeetr}
\bibliography{References_ISAC}
\end{document}

%% file: Preamble.tex
\usepackage{amsmath,epsfig}
\usepackage[T1]{fontenc}
\usepackage[utf8]{inputenc} 
\usepackage{psfrag}
\usepackage{array}
\usepackage{cite}
\usepackage{amsfonts}

\usepackage{amssymb}
\usepackage{color}
\usepackage[dvipsnames]{xcolor}
\usepackage[table]{xcolor}
\usepackage{rotating}
\usepackage{graphicx}
\usepackage{float}
\usepackage{circuitikz}
\usepackage{tikz}
\usetikzlibrary{arrows.meta, decorations.markings, shapes.multipart}

\usepackage{amsbsy} 
\usepackage{tcolorbox}
\usepackage{multirow}

\usepackage[nolist]{acronym}
\usepackage{doi}

\usepackage{bigfoot}

\usepackage{listings}
\usepackage{xcolor} 

\usepackage{colortbl}
\usepackage{adjustbox}

\makeatletter
\def\blfootnote{\xdef\@thefnmark{}\@footnotetext}
\makeatother

\usepackage{cuted}

\usepackage{booktabs}

\usepackage{subcaption}
\usepackage{caption}

\usepackage{stfloats} 

\usepackage{lipsum} 
\usepackage{mwe}    

\usepackage{amsthm}

\usepackage{tikz} 
\usetikzlibrary{arrows.meta, positioning, decorations.pathreplacing}

\usepackage{cuted}

\usepackage{pdflscape} 

\usepackage{rotating}

\usepackage{booktabs}
\usepackage{tabularx}

\usepackage{cleveref}
\crefname{table}{Table}{Tables}   
\Crefname{table}{Table}{Tables}   

\usepackage{stfloats}

\usepackage{siunitx}
\usepackage{geometry}

\usepackage{pdflscape}
\usepackage{fancyhdr} 
\usepackage{changepage}
\usepackage{cancel}
\usepackage{diagbox}

\fancypagestyle{mylandscape}{
\fancyhf{} 
\fancyfoot{
\makebox[\textwidth][r]{
  \rlap{\hspace{1.75cm}
    \smash{
      \raisebox{4.87in}{
        \rotatebox{90}{\thepage}}}}}}
}

\usepackage[section]{algorithm}
\usepackage{algorithmicx}
\usepackage{algpseudocode}

\usepackage{caption}
\definecolor{mirojo}{HTML}{cc3300}

\usepackage{booktabs}

\usepackage{fancyvrb} 
\VerbatimFootnotes    

\usepackage{geometry}

%% file: Acronyms.tex
\begin{acronym}
    \acro{3GPP}{Third Generation Partnership Project}
    \acro{5G}{fifth generation}
    \acro{6G}{sixth generation}
    \acro{BS}{base station}
    \acro{CRLB}{Cramér-Rao lower bound}
    \acro{CSI-RS}{channel state information reference signal}
    \acro{DDRS}{Doppler-delay reference signal}
    \acro{DFT}{discrete Fourier transform}
    \acro{DM-RS}{demodulation reference signal}
    \acro{FFT}{fast Fourier transform}
    \acro{IBFD}{in-band full duplex}
    \acro{ICI}{intercarrier interference}
    \acro{IDFT}{inverse \aclu{DFT}}
    \acro{IFFT}{inverse \aclu{FFT}}
    \acro{ISAC}{integrated sensing and communications}
    \acro{KPI}{key performance indicator}
    \acro{LLF}{log-likelihood function}
    \acro{ML}{maximum likelihood}    
    \acro{MSE}{mean squared error}
    \acro{NR}{new radio}
    \acro{OFDM}{orthogonal frequency-division multiplexing}
    \acro{PRS}{positioning reference signal}
    \acro{RAN}{radio access network}
    \acro{RB}{resource block}
    \acro{RCS}{radar cross section}
    \acro{RE}{resource element}
    \acro{RF}{radio frequency}
    \acro{RTT}{round-trip time}
    \acro{RV}{random variable}
    \acro{SCS}{subcarrier spacing}
    \acro{SNR}{signal-to-noise ratio}
    \acro{SRX}{sensing receiver}
    \acro{SS}{synchronization signal}
    \acro{ST}{sensing target}
    \acro{STX}{sensing transmitter}
    \acro{UAV}{unmanned aerial vehicle}
    \acro{UE}{user equipment}
\end{acronym}